\documentclass[letterpaper]{aa}

\usepackage{graphicx}
\usepackage{natbib}
\bibpunct{(}{)}{;}{a}{}{,}

\newcommand{\id}{\mbox{$\mathrm{d^{-1}}$}}
\newcommand{\Msun}{\mbox{$M_{\odot}$}}
\newcommand{\kms}{\mbox{$\mathrm{km\,s^{-1}}$}}

\newcommand{\Porb}{\mbox{$P_{\rm orb}$}}
\newcommand{\Pspin}{\mbox{$P_{\rm spin}$}}

\newcommand{\Line}[3]{\Ion{#1}{#2}\,$\lambda$\,#3}

\newcommand{\Ion}[2]{#1{\,\scriptsize #2}}
\newcommand{\Ha}{\mbox{${\mathrm H\alpha}$}}
\newcommand{\Hb}{\mbox{${\mathrm H\beta}$}}
\newcommand{\Hg}{\mbox{${\mathrm H\gamma}$}}

\begin{document}

\title{1RXS\,J062518.2+733433: A new intermediate polar}

\author{S. Araujo-Betancor\inst{1}\thanks{Visiting Astronomer, German-Spanish
Astronomical Center, Calar Alto, operated by the Max-Planck-Institut
f\"{u}r Astronomie, Heidelberg, jointly with the Spanish National
Commission for Astronomy.} \and
        B.T. G\"ansicke\inst{1}$^\star$
        H.-J. Hagen\inst{2}
        P. Rodriguez-Gil\inst{1} \and
        D. Engels\inst{2}
        }   
\offprints{S. Araujo-Betancor, e-mail: sab@astro.soton.ac.uk}

\institute{
Department of Physics and Astronomy, University of Southampton,
  Southampton SO17 1BJ, UK
\and
Hamburger Sternwarte, Universit\"at Hamburg, Gojenbergsweg 112,
21029 Hamburg, Germany}

\date{Received \underline{\hskip2cm} ; accepted \underline{\hskip2cm} }

\abstract{We report the identification of the cataclysmic variable
1RXS\,J062518.2+733433 as an intermediate polar. The orbital period of
the system is determined to be $283.0\pm0.2$\,min from the radial
velocity variation of \Ha, measured in an extensive set of
time-resolved spectroscopy. Differential optical photometry obtained
over a base line of three weeks reveals the presence of coherent
variability with a period of $19.788\pm0.002$\,min, which we suggest
to be the white dwarf spin period. The power spectrum of our
photometry also contains a strong signal near the spectroscopically
determined orbital period.  The emission lines in
1RXS\,J062518.2+733433 display a complex multicomponent structure. In
the trailed spectrogram of \Line{He}{I}{6678} we detect a narrow
component with a radial velocity semi-amplitude of $\simeq140$\,kms,
consistent with a possible origin on the irradiated face of the secondary. The
absence of eclipses gives an upper limit on the binary inclination of
$i\la60^\circ$.
\keywords{Stars: binaries: close -- Stars: individual:
  1RXS\,J062518.2+733433 -- Cataclysmic variables} } 

\maketitle

\section{Introduction}
Over the past few years it has become increasingly clear that our
understanding of the evolution of cataclysmic variables (CVs) is a
rather unsettled matter. A number of the predictions made by the
standard theory of CV evolution \cite[disrupted magnetic braking,
][]{king88-1} are in direct conflict with the observations.  Whereas a
number of additions/alternatives to the standard model have been
recently proposed \citep{king+schenker02-1, schenker+king02-1,
schenkeretal02-1, andronovetal03-1}, it is apparent that the
disturbing disagreements between the theory and the observations have
a common denominator: the possible impact of selection effects on the
currently known population of galactic CVs. In order to
quantitatively test any theory of CV evolution it is 
mandatory to establish a statistically complete sample of this class of binary stars.

There are currently a number of large scale surveys for CVs underway
that pursue this aim \citep[e.g.][]{szkodyetal02-2, marshetal02-1}.
Among these projects, our selection of CVs based on their
spectroscopic properties in the Hamburg Quasar Survey \citep[HQS;
][]{hagenetal95-1} has been especially prolific, resulting in the
discovery of more than 50 new bright CVs
\citep[e.g.][]{gaensickeetal00-2, nogamietal01-1, gaensickeetal02-2,
gaensickeetal02-3}.

In this paper, we report follow-up observations of
1RXS\,J062518.2+733433, henceforth RX\,J0625, an object originally
identified as a CV on the base of its X-ray emission
and optical spectrum \citep{weietal99-1}. We have
independently selected RX\,J0625 as a CV candidate because of the
noticeable Balmer emission in its HQS data. The strong
\Line{He}{II}{4686} emission detected in the identification spectrum
of RX\,J0625, along with coherent optical variability observed on a
time scale of $\sim20$\,min, immediately reveal the intermediate polar
nature of this CV. In Sect.\,2 we describe the observational data and
its reduction. We then derive the orbital period from the radial
velocity variation of the emission lines and the spin period from
differential photometry in Sect.\,3.  Finally, in Sect.\,4 we discuss
the behaviour of the strongest emission lines in detail and summarise
our findings in Sect.\,5.

\begin{figure}
\includegraphics[angle=-90,width=8.8cm]{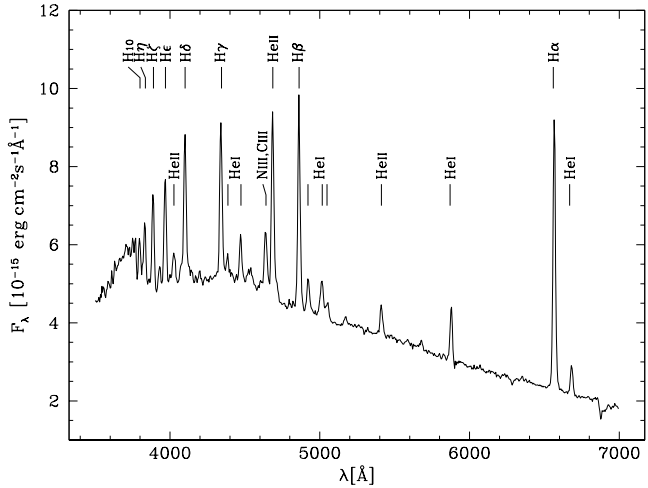}
\caption[]{\label{f-idspectrum} Identification spectrum of RX\,J0625
obtained on 2001 April 29.}
\end{figure}

\section{Observations and Data Reduction}
\subsection{Spectroscopy}
On 2001 April 29 we obtained a single identification spectrum of
RX\,J0625 at the Calar Alto 2.2m telescope  with the CAFOS focal
reductor spectrograph, using the standard SITe CCD (Table\,1). We used
the B-200 grism and a slit width of $2\arcsec$, resulting in a useful
wavelength range of 3500--7000\,\AA\ and a spectral resolution of
4.7\,\AA\ (Fig.\,\ref{f-idspectrum}).  A flux standard (BD+75$^\circ$\,325) was
observed with the same set up in order to correct for the instrumental
response. The identification spectrum contains noticeable Balmer,
\Ion{He}{I} and \Ion{He}{II} and N/C Bowen emission lines. The
strength of \Line{He}{II}{4686} is comparable to \Hb, indicating the
presence of a strong source of ionising photons in RX\,J0625, 
typical of either magnetic CVs or novalike variables. 
In order to determine the orbital period of RX\,J0625, we obtained 45
higher resolution spectra, again with CAFOS at the Calar Alto 2.2m telescope
(Table\,1).  This time the G-100 grism was used in conjuction with a
$1.2\arcsec$ slit which gave a wavelength range of 4240--8300\,\AA\ and
a spectral resolution of 2.1\,\AA. The higher resolution spectra were
obtained over a period of 3 weeks, optimising the sampling for an
efficient period determination.

All spectra were reduced in a standard manner using the
\texttt{Figaro} package within the Starlink software collection.  The
frames were corrected for the bias level by subtracting the mean of a
series of bias images taken at the start and end of each observing
night.  Dome flat-fields were used to remove pixel to pixel variations
of the chip.  The spectra were then optimally extracted
\citep{horne86-1} and sky line subtracted using Tom Marsh's
\texttt{Pamela} package. Especial care was taken to account for the
tilt of the spectra in order to maximise the signal-to-noise
ratio. The wavelength calibration was performed using mercury-cadmium,
helium-argon and rubidium arcs. Uncertainties on every data point
calculated from photon statistics are rigorously propagated through
every stage of the data reduction. We did not attempt to
flux-calibrate the higher resolution spectra as it was not required
for the radial velocity analysis described below.

Table~2 gives the equivalent widths and line widths of the main
emission lines detected in the average of the 45 high resolution
spectra.

\begin{table}[t]
\caption[]{\label{t-obslog}Log of Observations.}
\begin{flushleft}
\begin{tabular}{rcccc}
\hline\noalign{\smallskip}
Date & UT Time &  Data & Exp.(s) & Num. Obs \\  
\hline\noalign{\smallskip}
\multicolumn{5}{l}{Spectroscopy} \\
2001 Apr 29    &  20:37 - 20:47 & B-200  & 600 & 1 \\ 
2002 Dec 09    &  01:09 - 01:21 & G-100  & 600 & 2 \\ 
2002 Dec 13    &  04:05 - 05:44 & G-100  & 600 & 9 \\
2002 Dec 14    &  23:20 - 00:08 & G-100  & 600 & 5 \\ 
2002 Dec 15    &  03:00 - 03:45 & G-100  & 600 & 5 \\ 
2002 Dec 15    &  05:23 - 06:07 & G-100  & 600 & 5 \\ 
2002 Dec 15    &  22:47 - 23:32 & G-100  & 600 & 5 \\ 
2002 Dec 16    &  01:39 - 02:24 & G-100  & 600 & 5 \\
2002 Dec 28    &  23:04 - 23:26 & G-100  & 600 & 3 \\
2002 Dec 29    &  01:12 - 01:34 & G-100  & 600 & 3 \\
2002 Dec 29    &  03:58 - 04:20 & G-100  & 600 & 3 \\
\multicolumn{5}{l}{Photometry} \\
2002 Dec 09 & 01:29 - 06:23 &  V     & 30 & 386 \\ 
2002 Dec 15 & 00:57 - 02:14 &  Clear & 30 & 100 \\
2002 Dec 15 & 02:33 - 02:39 &  Clear & 30 &  10 \\
2002 Dec 16 & 00:00 - 01:15 &  Clear & 30 & 100 \\
2002 Dec 28 & 21:07 - 22:23 &  V     & 30 & 100 \\
2002 Dec 28 & 23:43 - 00:51 &  V     & 30 &  98 \\
2002 Dec 29 & 01:48 - 03:41 &  V     & 30 & 270 \\
2002 Dec 29 & 23:55 - 03:36 &  V     & 30 &  23 \\
2002 Dec 30 & 03:55 - 04:48 &  V     & 30 & 293 \\
2002 Dec 31 & 04:11 - 06:00 &  V     & 30 & 136 \\
\noalign{\smallskip}\hline
\end{tabular}
\end{flushleft}
\end{table}

\begin{table}[t]
\caption[]{\label{t-ew}Equivalent widths and line widths (corrected
for the instrumental resolution) of the strongest emission lines
measured from the average of the 45 spectra obtained in December
2002.}
\begin{flushleft}
\begin{tabular}{lrr}
\hline\noalign{\smallskip}
Line & FWHM [\AA] & EW [\AA] \\
\hline\noalign{\smallskip}
\Hg  & $13.7\pm0.2$ & $14.3\pm1.0$  \\
\Line{He}{II}{4686} & $14.2\pm0.3$ & $15.0\pm2.0$ \\
\Hb  & $14.2\pm0.2$ & $22.0\pm2.0$  \\
\Line{He}{I}{5876}  & $14.0\pm0.3$ & $ 9.6\pm0.5$ \\
\Ha  & $15.8\pm0.2$ & $64.0\pm2.0$  \\
\Line{He}{I}{6678}  & $17.3\pm0.2$ & $ 7.5\pm0.5$ \\
\noalign{\smallskip}\hline
\end{tabular}
\end{flushleft}
\rmfamily
\end{table}

\subsection{Photometry}
\begin{figure}
\includegraphics[width=8.8cm]{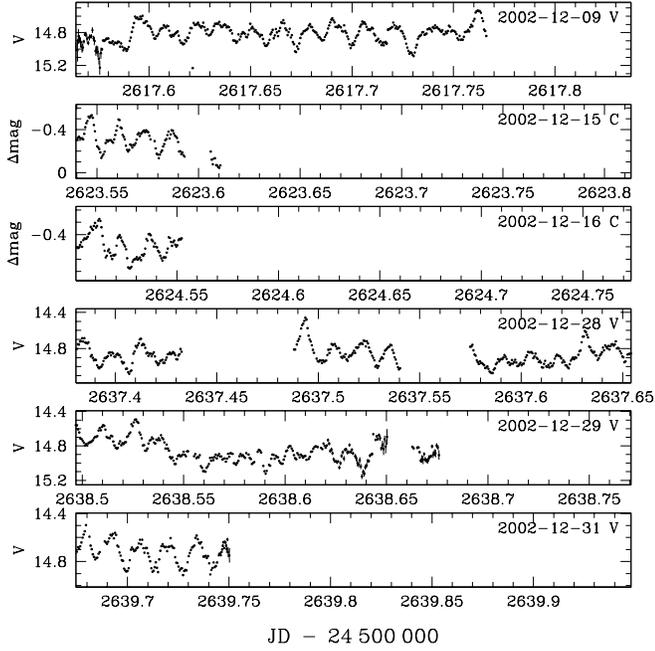}
\caption[]{\label{f-lc_all} Differential CCD $V$-band (V) and
filter-less (C) photometry obtained at
the Calar Alto 2.2m observatory. Note that the different
scales of the filter-less and $V$ band data.}
\end{figure}

We have obtained differential $V$-band and filter-less photometry of
RX\,J0625 during 6 nights in December 2002 with the CAFOS SITe CCD
camera on the Calar Alto 2.2m telescope (Table\,\ref{t-obslog}). In
order to achieve a high time resolution ($\sim30$\,s), only a small
window of the chip ($\sim3\arcmin\times 2\arcmin$) was read out. The
data were bias-subtracted and flat-fielded in a standard fashion using
the \texttt{ESO-MIDAS} package, and aperture magnitudes were extracted
with the sextractor \citep{bertin+arnouts96-1}. The $V$ magnitudes of
RX\,J0625 were derived relative to the HST Guide Star GSC\,0437000234
($V=13.4\pm0.4$), located $1\arcmin50\arcsec$ southwards of RX\,J0625.
This comparison star has been saturated in a number of the filter-less
CCD images obtained on December 14/15, and, hence, we used
GSC\,0437000998 ($V=14.9\pm0.4$), located $1\arcmin10\arcsec$
northwards of RX\,J0625 for the reduction of these
images. RX\,J0625 was found at an average magnitude of
$14.80\pm0.05$ throughout the nights in which we used the $V$ band
filter. The light curves of RX\,J0625 clearly reveal the presence of
variability with a period of $\sim20$\,min and an amplitude of
$\sim\pm0.2$\,mag throughout all nights (Fig.\,\ref{f-lc_all}). In
addition to this short-term variations the light curve of RX\,J0625
displays a modulation on time scales of several hours.

\begin{figure}
\centerline{\includegraphics[width=5.2cm]{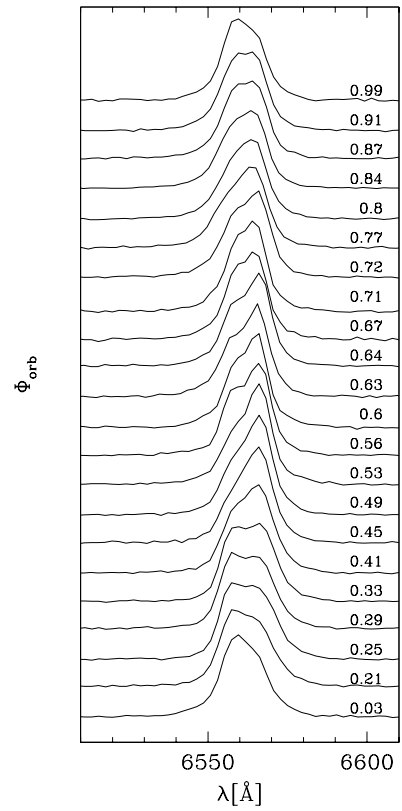}}
\caption[]{\label{f-halpha} Continuum-subtracted and
normalised \Ha\ profile sample sorted by  orbital phases computed
from Eq.\,(\ref{e-orb_ephemeris}).}
\end{figure}

\begin{figure}
\includegraphics[width=8.8cm]{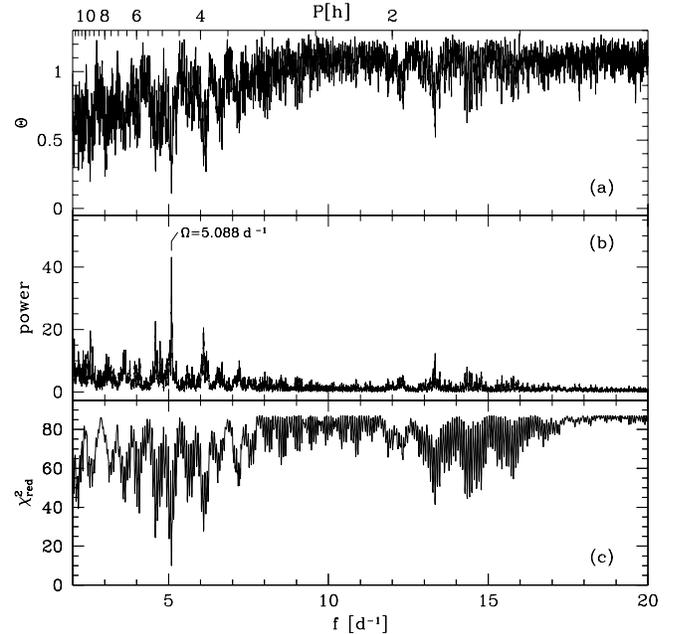}
\caption[]{\label{f-rvperiod} Period analysis of the radial velocities
measured from a Gauss-fit to the \Ha\ emission line profiles. (a):
Phase dispersion method periodogram. (b): Analysis of variance
periodogram. (c): $\chi^2$ sine fit.}
\end{figure}

\section{Analysis}
\subsection{\label{s-radvel}Spectroscopy}
\begin{figure}
\includegraphics[angle=-90,width=8.8cm]{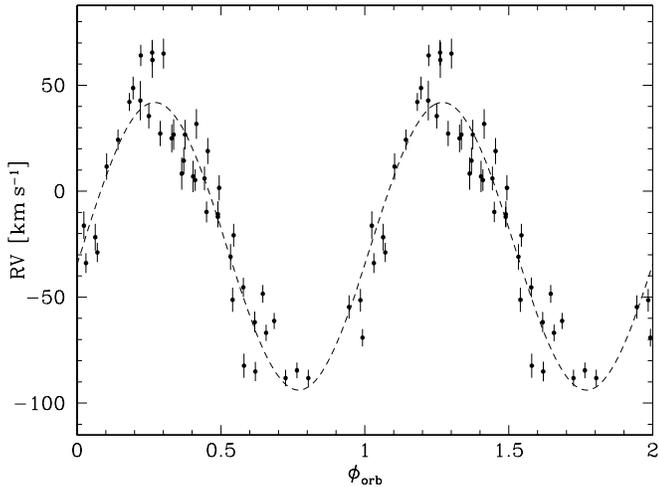}
\caption[]{\label{f-rvfold} Radial velocities measured from a
Gauss-fit to the \Ha\ emission line profiles, folded with the orbital period
$\Porb=283.0$\,min, using Eq.\,(\ref{e-orb_ephemeris}). The dashed
curve is the best sine fit to the radial velocities.}
\end{figure}

The primary aim of our time resolved spectroscopy of RX\,J0625 is to
measure its orbital period from the radial velocity variations of the
emission lines. The line profiles clearly display a complicated
multicomponent structure (Fig.\,\ref{f-halpha}). In a first attempt,
we used a single Gaussian least square fitting procedure to determine
the radial velocity variations of \Ha, \Hb\ and
\Line{He}{II}{4686}. The radial velocity data were then subjected to
the following period analysis methods: (a) the phase dispersion method
\citep{stellingwerf75-1}, (b) the analysis of variance \citep[][ as
implemented in the \texttt{MIDAS} context \texttt{TSA}]{schwarzenberg-czerny89-1}, (c)
and a sine wave $\chi^2$ fitting procedure, using \texttt{chisq} in
the Starlink package \texttt{period}. The resulting periodograms for
\Ha\ are shown in Fig.\,\ref{f-rvperiod}.  The orbital frequency
(period) is inferred from the three independent techniques in a
consistent way to be $\Omega=5.088\pm0.003$\,\id
($\Porb=283.0\pm0.2$\,min), where the error is computed from the
$\chi^2$ sine wave fitting ($2\sigma$, 4 interesting parameter). A
slightly more conservative error estimate based on the results
obtained from all three methods gives $\Porb=283.0\pm0.3$\,min. The
analysis of \Hb\ and \Line{He}{II}{4686} provides consistent results
for \Porb, but with significantly larger errors. The robustness of our
period determination is shown in Fig.\,\ref{f-rvfold} were we folded
the \Ha\ radial velocities over the orbital period. The radial
velocity measurements from this fit are provided in Table\,3.

\begin{table*}
\caption[]{\label{t-rv} Radial velocities of the \Ha\ emission line determined from
a single-Gaussian fit to the line profile (corrected for the
heliocentric velocity).}
\begin{tabular}{lllllllll}
\hline\noalign{\smallskip}
HJD & RV & RV error & HJD & RV & RV error & HJD & RV & RV error \\
--2450000 & [\kms] & [\kms] & --2450000 & [\kms] & [\kms] & --2450000 & [\kms] & [\kms] \\
\hline\noalign{\smallskip}
617.5515 &  42.7 &  9.3 &    623.5093 & -31.0 &   6.0  &   624.4842 &   1.6 &  6.0 \\
617.5598 &  61.9 &  8.5 &    623.6290 &  24.3 &   4.8  &   624.5729 & -54.7 &  5.5 \\
621.6740 &  48.7 &  5.3 &    623.6368 &  42.1 &   4.2  &   624.5807 & -51.5 &  5.4 \\
621.6847 &  35.4 &  5.9 &    623.6446 &  64.0 &   5.0  &   624.5884 & -16.3 &  6.4 \\
621.6925 &  27.2 &  6.0 &    623.6524 &  65.4 &   6.0  &   624.5962 & -21.8 &  6.3 \\
621.7003 &  24.9 &  6.5 &    623.6601 &  64.9 &   7.1  &   624.6040 &  11.6 &  6.3 \\
621.7085 &  14.4 &  6.2 &    623.7279 & -48.5 &   4.2  &   637.4651 & -51.3 &  5.7 \\
621.7162 &   5.2 &  5.1 &    623.7357 & -61.2 &   3.8  &   637.4729 & -82.4 &  5.7 \\
621.7240 &  -9.8 &  4.8 &    623.7434 & -88.2 &   3.9  &   637.4807 & -85.1 &  4.5 \\
621.7318 & -12.1 &  4.6 &    623.7512 & -84.6 &   3.8  &   637.5538 & -69.1 &  4.1 \\
621.7425 & -20.8 &  5.4 &    623.7590 & -88.2 &   4.0  &   637.5616 & -33.9 &  4.7 \\
623.4760 &   8.3 &  7.6 &    624.4531 &  26.7 &   7.2  &   637.5694 & -28.9 &  4.5 \\
623.4838 &   6.9 &  7.4 &    624.4609 &  26.7 &   7.0  &   637.6690 & -45.4 &  4.5 \\
623.4915 &   6.0 &  5.6 &    624.4687 &  31.7 &   7.0  &   637.6768 & -61.9 &  4.7 \\
623.5007 & -10.9 &  6.1 &    624.4764 &  18.9 &   6.2  &   637.6846 & -66.9 &  3.9 \\
\hline\noalign{\smallskip}
\end{tabular}
\end{table*}

For comparison purposes we measured the radial velocities of \Ha\
cross-correlating the observed line profile with a single Gaussian of
fixed width of 300\,\kms\ \citep{schneider+young80-2}.
The resulting orbital period is identical to that derived from the
single Gaussian fit to the \Ha\ profiles, but the radial velocity
variation displays an amplitude larger by a factor of $\sim2$.

Finally, we attempted to fit the \Ha\ emission with a blend of a
narrow and a broad Gaussian component to model more accurately the
complex structure of the observed profiles. Unfortunately, this
procedure did not provide unambiguous results because of the limited
spectral resolution of our data. Due to the multicomponent structure
of the lines profile (see Fig.\,\ref{f-halpha}) it is difficult to
interpret the radial velocity variation obtained from either the
single Gaussian fit (Fig.\,\ref{f-rvfold}) or the Gauss correlation to
\Ha. The behaviour of the most important emission lines is discussed
in more detail in Sect.\,\ref{s-discussion} on the basis of
trailed spectrograms.

\subsection{Photometry}
\begin{figure*}
\includegraphics[angle=-90,width=18cm]{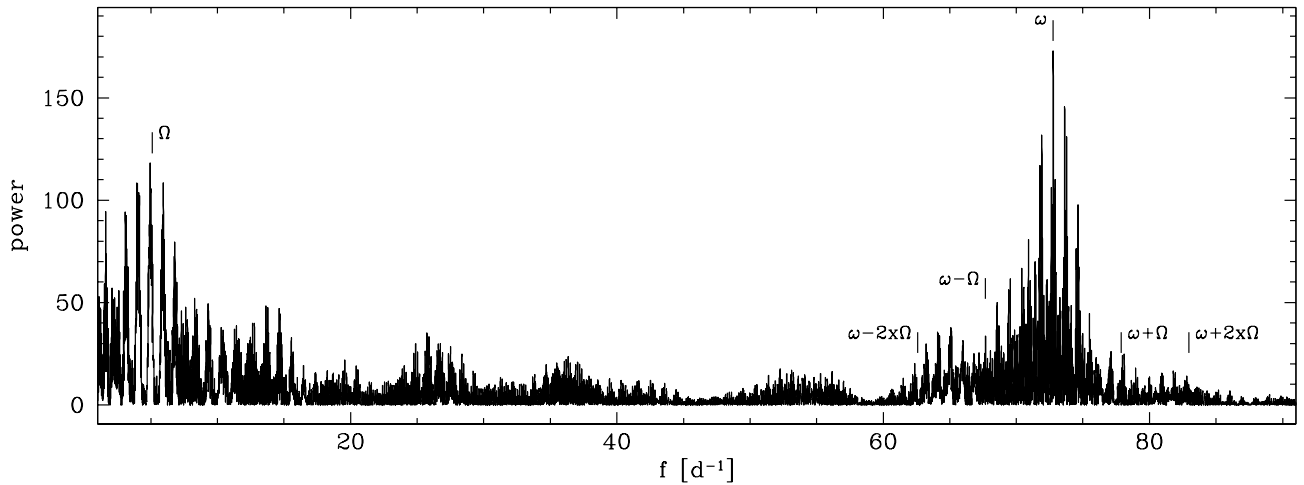}
\caption[]{\label{f-all_power} Scargle periodogram of the differential
CCD photometry shown in Fig.\,\ref{f-lc_all}. $\omega$ and $\Omega$
are the white dwarf spin and orbital frequency, respectively.}
\end{figure*}

\begin{figure}
\includegraphics[angle=-90,width=8.8cm]{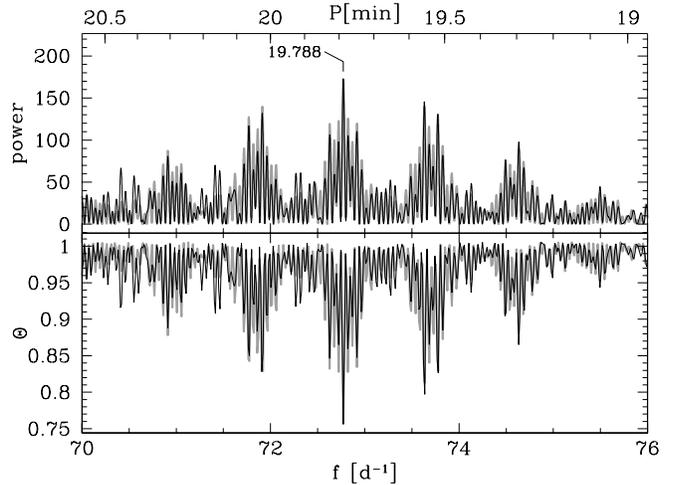}
\caption[]{\label{f-spin_power} Period analysis near the presumed
white dwarf spin. Top panel: Scargle periodogram.  Bottom panel: phase
dispersion periodogram. The black lines are computed from the observed
data. The gray lines are computed from a sinusoid assuming a period of
$1/\omega=19.788$\,min and identical sampling as in the observed
data. }
\end{figure}

In order to analyse the periodicities observed in the light curve of
RX\,J0625 (Fig.\,\ref{f-lc_all}), we have computed from the entire
photometric data set both a \citet{scargle82-1} periodogram (as
implemented in the \texttt{MIDAS} context \texttt{TSA}) as well as a periodogram using
the Phase Dispersion Method (PDM) of \citet{stellingwerf75-1}.
Because the lengths of our individual photometric observations are of
the order of or shorter than the orbital period subtracting the
nightly mean from the data would introduce erroneous signals in the
low-frequency range of the periodograms. Considering that RX\,J0625
does not exhibit noticeable night-to-night variability, we therefore
decided to subtract from the $V$ filter and white light data the mean
of all measurements obtained in the corresponding band prior to the
computation of the periodograms. The resulting periodograms
(Fig.\,\ref{f-all_power} and \ref{f-spin_power}) contain strong
signals concentrated in the frequency ranges $f\la10\,\id$ and
$f\simeq60-70\,\id$. The nature of the signals detected in these two
separate frequency regimes is discussed below.

\paragraph{High-frequency signal.}
Figure\,\ref{f-spin_power} shows an enlargement around the peak signal
contained in the high-frequency range of the periodogram shown
in Fig.\,\ref{f-all_power}. Both period analysis methods consistently
detect the maximum power at $72.772\pm0.008\,\id$,
corresponding to a period of $19.788\pm0.002$\,min,
whose error is given by $1\sigma$ width of the Gaussian fit to
the frequency peak. In order to test the significance of this signal,
we have created a faked data set from a sine wave with a period of
19.788\,min, a similar amplitude as the observed short-period
variability, and sampled exactly in the same way as the observed
data. In addition, we added a 10\% error to these synthetic data. The
periodograms for this faked data set are plotted in gray in
Fig.\,\ref{f-spin_power}, and reproduce very well the entire structure
of aliases contained in the periodograms of the observed data. We
conclude from this comparison that the frequency of the maximum signal
detected in the periodograms is indeed the true period underlying the
short-term variability observed in RX\,J0625.

\begin{figure}
\includegraphics[width=8.8cm]{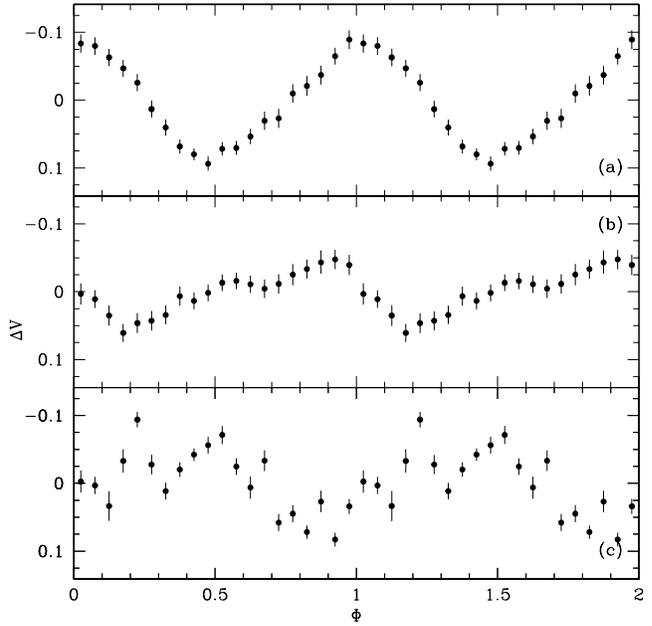}
\caption[]{\label{f-folded} The photometric data points from
Fig.\,\ref{f-lc_all} mean-subtracted and folded over: (a) The spin
period of 19.788\,min, using Eq.\,(\ref{e-spin_ephemeris}). (b) The
beat period of 21.275\,min. (c) The orbital period of 283.0\,min,
using Eq.\,(\ref{e-orb_ephemeris}).}
\end{figure}

Coherent variability on time scales of a several minutes to several
tens of minutes has been detected in the optical light curves of a
number of intermediate polars \citep[e.g.][]{patterson94-1}, and is
interpreted as the spin period of the accreting magnetic white
dwarf. We conclude from the spectral appearance of RX\,J0625 and from
the detected coherent optical variability that RX\,J0625 is indeed a
new member of the small class of intermediate polars.  We
suggest that the detected frequency (period) of 72.772\,\id\
(19.788\,min) is the white dwarf spin frequency $\omega$ (period
$\Pspin=1/\omega$). From the combined $V$ band and filter-less
photometry, we derive the following spin ephemeris:
\begin{equation}
\label{e-spin_ephemeris}
\phi_0^\mathrm{spin} = \mathrm{HJD}\,2452617.0645(4) + 0.013742(3)\times E
\end{equation}
where $\phi^\mathrm{spin}=0$ is defined as the spin pulse
maximum. Errors in the last digit are given in
brackets. Figure\,\ref{f-folded}(a) shows our photometric data folded
with this ephemeris. The quasi-sinusoidal shape of this light
curve, which is typical for the spin light curve of intermediate
polars (see, e.g., the spin light curve of FO\,Aqr by
\citealt{demartinoetal94-1}), lends support to our hypothesis that the
72.772\,\id\ frequency detected in the power spectrum of RX\,J0625 is
indeed the spin frequency of the white dwarf.

Another hallmark of intermediate polars is the detection of beat
frequencies between the white dwarf spin frequency $\omega$ and the
orbital frequency $\Omega$, which arise from the reprocession of
X-rays emitted from close to white dwarf surface on, e.g., the
secondary star. Such sideband signals have been detected in different
systems at $\omega-2\Omega$, $\omega-\Omega$, $\omega+\Omega$,
$\omega+2\Omega$ (see \citealt{warner86-2} for an interpretation of
these frequencies). In RX\,J0625, the only unambiguous detection of a
sideband signal is $\omega-\Omega$ (see Fig.\,\ref{f-all_power}). The
photometric data folded over this beat period of $21.275$\,min
are shown in Fig.\,\ref{f-folded}(b). While we are
confident that our identifications of the spin and beat period are
correct we mention as a note of caution that time-resolved X-ray
and/or polarimetric data are necessary to unambigously confirm this
interpretation.

\paragraph{Low-frequency signal.} 
The strongest signal in the low-frequency range does not coincide with
the orbital frequency derived in Sect.\,\ref{s-radvel} from the radial
velocity variation of \Ha, but is found at 4.944\,\id, corresponding
to a period of 291.3\,min. This period is $\simeq3\%$ longer than the
orbital period. Considering the sampling of our photometric
time series, especially the fact that none of our data sets covers
significantly more than one binary orbit, the most likely hypothesis
is that the detected low-frequency signal is related to the orbital
motion of the binary. Further observations would be useful to test in
detail whether the low-frequency photometric frequency is indeed
identical to the orbital frequency $\Omega$.

\section{Discussion\label{s-discussion}}
\begin{figure*}
\includegraphics[angle=-90,width=18cm]{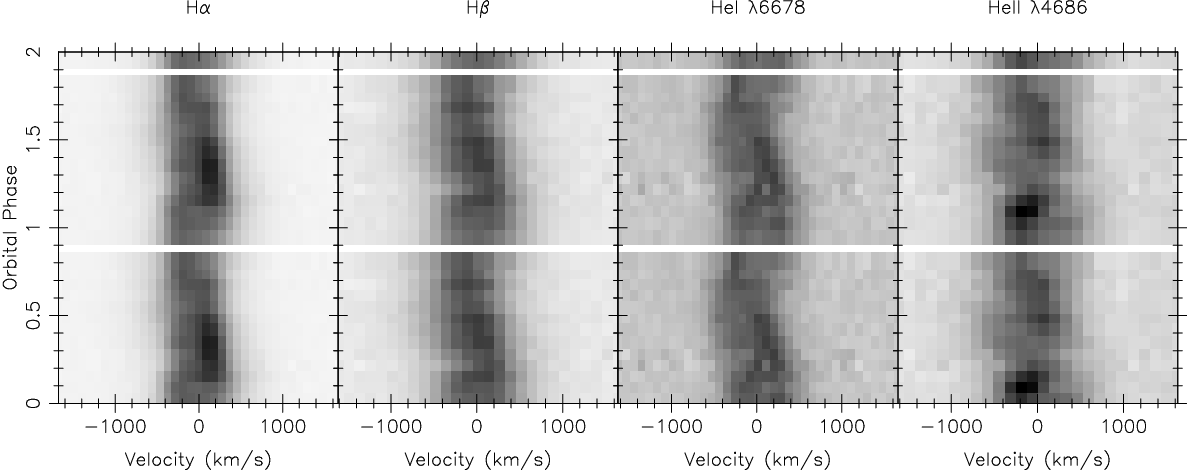}
\caption[]{\label{f-trailed_spec} Trailed spectrograms of \Ha, \Hb,
\Line{He}{I}{6678}, \Line{He}{II}{4686}, computed from the 45 spectra
obtained in December 2002 (Table\,1). Orbital phases have been
computed using Eq.\,(\ref{e-orb_ephemeris}).}
\end{figure*}

As already mentioned in Sect.\,\ref{s-radvel}, the line profiles
clearly display a multicomponent structure (Fig.\,\ref{f-halpha}),
which makes the interpretation of the determined radial velocity
variations rather ambiguous (Fig.\,\ref{f-rvfold}). To explore the
behaviour of the emission lines in more detail, we constructed trailed
spectra of \Ha, \Hb, \Line{He}{I}{6678}, and \Line{He}{II}{4686},
which are shown in Figure\,\ref{f-trailed_spec}. These diagrams were
computed from the continuum-normalised spectra after binning into 15
phase intervals. The \Line{He}{I}{6678} line clearly shows two
emission components: a narrow one with a radial velocity
semi-amplitude of $\approx 140$\,\kms\ and a wider one with velocity
reaching $\sim \pm 500$\,\kms. Both components are not in phase, the
phase offset being $\sim 0.15$.  The trailed spectra shown in
Fig.\,\ref{f-trailed_spec} were phased according to the following
orbital ephemeris
\begin{equation}
\label{e-orb_ephemeris}
\phi_0^\mathrm{orb} = \mathrm{HJD}\,2452617.5083(9) + 0.1965(1)\times E
\end{equation}
where the instant of zero phase was derived as the time of blue-to-red
crossing of the narrow component of \Line{He}{I}{6678}.

In analogy to the narrow emission lines observed in many magnetic CVs,
this narrow emission line component may have its origin on the
irradiated face of the secondary star. In order to test this
hypothesis, we performed the following simple calculation.  The
secondary mass was estimated from the mass-period relation derived by
\citet{smith+dhillon98-1}, which gives a value of $M_2 \approx
0.48$\,\Msun. For the primary mass we adopted $M_1=0.8$\,\Msun\
\citep[the average white dwarf mass in long-period CVs determined
by][]{smith+dhillon98-1}, resulting in a mass ratio $q=M_2/M_1 \approx
0.6$. Assuming a disc radius of $0.8\,R_\mathrm{L_1}$, the lack of
eclipses in the optical light curve of RX\,J0625 sets an upper limit
on the orbital inclination of $i \la 60^\mathrm{o}$. The projected
velocity of the secondary is given by:
\begin{equation}
K_2=\frac{2 \pi a \sin i}{P_\mathrm{orb} (1+q)}~,
\end{equation}   
\noindent
where $a$ is the binary separation. With the above assumptions we the
radial velocity variation of the secondary star can be as large as
$K_2 \approx 130$\,\kms, which is comparable to the value obtained
from the trailed spectrogram of the narrow \Line{He}{I}{6678}
line. The predicted $K_1$ velocity is then $\approx 80$\,\kms, somewhat
larger than the semi-amplitude of the radial velocity curve of the
broad component of \Ha\ (Fig.\,\ref{f-trailed_spec}, see also
Fig.\,\ref{f-rvfold}), but not inconsistent. 

Taken at face value, the behaviour of the narrow emission line
components detected in our (admittedly rather low resolution)
phase-resolved spectroscopy are consistent with an origin on the
irradiated face of the secondary.

As we mentioned above, the broad component of the \Line{He}{I}{6678}
line displays a phase offset of $\sim 0.15$ with respect to the narrow
one, reaching a velocity of $\sim -500$\,\kms\ at its maximum
excursion to the blue. The phasing with respect to the narrow
component (and assuming that our above interpretation of the origin of
the narrow component is correct) indicates that the broad emission is
not coming from an axisymmetric structure around the white
dwarf. Nevertheless, the high velocity dispersion observed in the
broad component clearly points towards an origin close to the white
dwarf. The maximum velocity to the blue is reached at phase $\sim
0.4-0.5$. A similar high-velocity S-wave has been observed in the
intermediate polars EX\,Hya \citep{hellieretal89-1} and V1025\,Cen
\citep{hellieretal02-1}. The same components are also characteristic
of the SW\,Sextantis stars, which have been recently proposed to be
magnetic systems \citep{rodriguez-giletal01-1}. In all these systems,
it is believed that these high-velocity S-waves form in the vicinity
of the primary's magnetosphere.
 
The trailed spectra of \Line{He}{II}{4686} seem to exhibit the same
two emission components, with the narrow one also dominating and same
velocity amplitudes as in \Line{He}{I}{6678}. Also the trailed spectra
of \Ha\ and \Hb\ clearly show a multicomponent structure, even though
their narrow components are less obvious than in \Line{He}{I}{6678}. In
order to securely identify the different line components with physical
emission sites in the binary, it will be necessary to obtain
time-resolved spectroscopy study spanning several consecutive binary
orbits with better spectral resolution.

\section{Conclusions}
We have observed RX\,J0625 as part of our ongoing search for new CVs
selected on the base of their spectroscopic properties from the
Hamburg Quasar Survey. From the radial velocity variations measured in
\Ha, we determine an orbital period of $283.0\pm0.2$\,min. The
detection of coherent optical variations clearly classifies RX\,J0625
as being a member of the small class of intermediate polars.  The
period of these variations, $19.788\pm0.002$\,min, which is
most likely the spin period of the white dwarf. In terms of its
orbital and spin period, RX\,J0625 is very similar to FO\,Aqr
($\Porb=291$\,min, and $\Pspin=21$\,min). Our phase-resolved
spectroscopy clearly shows that the emission lines are multicomponent
structured, and we identify a narrow component with a radial velocity
semi-amplitude of $\simeq140$\,kms, which might originate on the
irradiated face of the secondary. Extensive photometric monitoring,
phase-resolved high-resolution spectroscopy and pointed X-ray
observations of RX\,J0625 are strongly encouraged.

\acknowledgements SAB likes to thank PPARC for a studentship. BTG was
supported by a PPARC Advanced Fellowship, The HQS was supported by the
Deutsche Forschungsgemeinschaft through grants Re\,353/11 and
Re\,353/22. We thank Ana Guijarro for obtaining part of the Calar Alto
observations. We are grateful to Christian Knigge for suggesting the
use of \texttt{chisq} in the Starlink \texttt{period} package for the
analysis of the radial velocity data~---~and for pointing out that the
latest version of \texttt{period} does not work properly. Tom Marsh is
thanked for providing his reduction and analysis packages. Patrick 
Woudt is acknowldedged for his prompt referee report and for a number
of useful suggestions.

Note added in proof: After submission of this paper, we became
aware of an independent photometric study of RX\,J0625 by
\citet{staudeetal03-1} that confirms the likely white dwarf spin
period and shows that the low-frequency photometric signal detected in
RX\,J0625 is consistent with our spectroscopic orbital period within
the errors of the measurements. Furthermore, the referee drew our
attention to the fact that an orbital period for RX\,J0625 has been
included on December 13 2002 in the online catalogue of
\citet{downesetal01-1}, which is consistent with our value.

\bibliographystyle{aa}

\end{document}